\documentclass[secnumarabic, prd]{revtex4}
\usepackage{bm}

\begin{document}

\title{Relativity from absoluteness}
\author{Valery P. Dmitriyev}
\affiliation{Lomonosov University, \\
Moscow, Russia}
 \email{aether@yandex.ru}
\date{\today}

\begin{abstract}
The shortening of bodies in the direction of motion, Lorentz contraction, follows from the solution of Maxwell's equations. Moving light clocks will tick slower than those at rest because the speed of light does not depend on a source of the light. The latter and  Lorentz contraction imply the relativistic time dilation. The invariance of the light speed defined as the round-trip value follows from the time dilation and Lorentz contraction. An observer is incognizant about his motion relative to the absolute frame of reference. So, in order to synchronize spaced clocks in a moving reference frame he uses the same procedure as in the absolute frame. We deduce the Lorentz transformation from the Lorentz contraction, time dilation, invariance of the light speed and synchronization procedure. Lorentz transformations  constitute a symmetry group of Maxwell's equations.  That is the reason why the absolute frame can not be distinguished among other inertial reference frames.
\end{abstract}


\maketitle

\section{Basic assumptions}

Special relativity theory (SRT) shows how measurements of the distance and time in one inertial frame of reference are related with measurements of the distance and time in other inertial  reference frame.  Inertial reference frames (IRF) are defined as reference frames that move uniformly along a straight line relative to each other.

The absolutist approach to SRT rests on following assumptions.

\emph{There is a reference frame where Maxwell's equations hold in
their usual form.} We \emph{interpret} it as the \emph{absolute reference frame} (ARF).

\emph{The speed of light does not depend on the motion of the light's
source.}

In the present report we will derive transformations of the one space coordinate and time.  In this derivation we will essentially do without any additional space coordinate.

\section{Lorentz contraction is a real physical effect}

Let we be in ARF where Maxwell's equations are valid. We are interested in the interaction of two electric charges. We have for the scalar potential $\varphi_0$ of the charge $q$  that is at rest in ARF
\begin{equation}
\partial_x^2\varphi_0 = -4\pi q\delta(x). \label{1}
\end{equation}
\label{note}(A note. Here and further on equations in partial derivatives are conventionally
written down in an abridged form: in order not to overload the report by not so essential
 details, components in $y$ and $z$ are dropped.) We will choose the Lorenz gauge for the sake of convenience. The scalar potential $\varphi$ of a charge that moves uniformly in ARF with the velocity ${v}$ obeys the equation
\begin{equation}
\partial_x^2\varphi - c^{-2}\partial_t^2\varphi = -4\pi q\delta(x - {v}t). \label{2}
\end{equation}
We will express $\varphi$ via $\varphi_0$. Notice that the electric field moves
together with the charge. Therefore the function  $\varphi$ must have the form
\begin{equation}
\varphi(x - {v}t). \label{3}
\end{equation}
We will pass to the moving RF by means of the transformation of the space coordinate
\begin{equation}
x^* = x - {v}t. \label{4}
\end{equation}
Differentiating (\ref{3}) with the account of (\ref{4}), $\partial_x^2\varphi = \partial_{x^*}^2\varphi$, $\partial_t^2\varphi = {v}^2\partial_{x^*}^2\varphi$,  and substituting it in (\ref{2}) we get
\begin{equation}
\gamma^{-2}\partial_{x^*}^2\varphi = -4\pi q\delta(x^*) \label{5}
\end{equation}
where
\begin{equation}
\gamma = \frac{1}{\sqrt{1 - {v}^2/c^2}}. \label{6}
\end{equation}
We will pass to the scaled coordinate
\begin{equation}
x' = \gamma x^*. \label{7}
\end{equation}
Substituting (\ref{7}) in (\ref{5}) and using the property of the $\delta$-function
\begin{equation}
\delta(ax) = \delta(x)/a \label{8}
\end{equation}
we get
\begin{equation}
\partial_{x'}^2\varphi = -\gamma 4\pi q\delta(x'). \label{9}
\end{equation}
Comparing equations (\ref{9}) and (\ref{1}) we find with the account of (\ref{7}) and (\ref{4})
\begin{equation}
\varphi = \gamma\varphi_0(\gamma(x - {v}t)). \label{10}
\end{equation}
We have for vector potential $A$ of the moving charge
\begin{equation}
\partial_x^2 A - c^{-2}\partial_t^2 A = -4\pi q\delta(x - {v}t){v}/c. \label{11}
\end{equation}
Comparing equations (\ref{2}) and (\ref{11}) we see that
\begin{equation}
A = \varphi{v}/c. \label{12}
\end{equation}
Substituting (\ref{10}) in (\ref{12}):
\begin{equation}
A=\gamma\varphi_0(\gamma(x - {v}t)){v}/c. \label{13}
\end{equation}
By definiton, the electric field is
\begin{equation}
E = -\partial_x\varphi - \partial_t A/c. \label{14}
\end{equation}
Substituting (\ref{10}) and (\ref{13}) in (\ref{14}):
\begin{equation}
E = -\gamma[\partial_x + ({v}/c^2)\partial_t]\varphi_0
(\gamma(x - {v}t)). \label{15}
\end{equation}
Using (\ref{4}) and (\ref{7}) in (\ref{15}) we get with the account of (\ref{6})
\begin{equation}
E = -\gamma^2[\partial_{x'} - ({v}^2/c^2)\partial_{x'}]\varphi_0(x') =
-\partial_{x'}\varphi_0(x'). \label{16}
\end{equation}
Comparing (\ref{16}) with the expression for the electric field of the charge at rest
\begin{equation}
E_0 = -\partial_x\varphi_0(x) \label{17}
\end{equation}
we find finally with the account of (\ref{4}) and (\ref{7})
\begin{equation}
E = E_0(\gamma(x - {v}t)). \label{18}
\end{equation}
In one dimension the magnetic component is vanishing. Thereby the interaction  reduces to the electric one. By (\ref{18}), the force field contracts $\gamma$ times in the direction of motion. We will come to the same conclusion if we will consider the field of a moving charge in three dimensions \cite{Dmitriyev}. The form (\ref{18}) enables us to assume that any body which moves uniformly with the velocity  ${v}$  relative to ARF contracts $\gamma$ times in the direction of the motion
\begin{equation}
l = l_0/\gamma \label{19}
\end{equation}
where $l_0$ is the length of a body at rest and $l$ the length of
the body in motion (see Appendix \ref{contraction}).

\section{The distance is the reading of a measuring device, the
foot-rule}

When a meter stick is moving its length decreases in accord with the
relation (\ref{19}). We will denote by $\Delta x'$ the distance between the two uniformly moving bodies as measured in the moving reference frame, and by  $\Delta x$ the same distance measured in ARF. The procedure of measuring the distance by a meter stick implies
\begin{equation}
l\Delta x' = l_0\Delta x. \label{20}
\end{equation}
Substituting (\ref{19}) into the left-hand part of (\ref{20}) we find
\begin{equation}
\Delta x' = \gamma\Delta x. \label{21}
\end{equation}
Next, let one of the bodies is in the origin of the moving reference frame. Then the coordinate of the other body is
\begin{equation}
x' = \Delta x'. \label{22}
\end{equation}
If origins of IRF and ARF coincide at a zeroth moment of time then by the Galileo transformation
\begin{equation}
\Delta x = x - {v}t. \label{23}
\end{equation}
Substituting (\ref{22}) into the left-hand part of (\ref{21}), and (\ref{23}) into the right-hand part of (\ref{21}) we find the transformation of the space coordinate to the moving reference frame
\begin{equation}
x' = \gamma(x - {v}t). \label{24}
\end{equation}
Formula (\ref{24}) agrees with the argument of the function (\ref{18}) rendering a moving force field.

\section{Dilation of time}

The second basic assumption (see $\S 1$) leads to that a moving clock would show the time different from the clock at rest. While the  Lorentz contraction makes an amendment to this inference.

The clock-work is always based on some periodic (oscillatory) process.  The period of
 vibrations can be taken as a unit of time (duration). To measure
the time we commonly use a pendulum (see Appendix \ref{pendulum}). In the current context the clock is meant to be set in fast translatory motion. But we do not yet know
 the equation of high speed mechanics (see Appendix \ref{mechanics}). Therefore we will
  use a device, known as a light clock, whose law of action was taken by us as one of the
  postulates: the speed of light does not depend on motion of a source.

A pulse of light circulates between two mirrors. The mirrors are mounted on a rigid bed one against another at a distance $l_0$. When the light clock is at rest in ARF, the time that the light propagates there and back, i.e. the period of the process, is
\begin{equation}
\tau_0 = 2l_0/c. \label{25}
\end{equation}
Denoting by $l$ the distance between the mirrors of the moving clock we have for the path in the forward direction
\begin{equation}
c\tau_+ = l + {v}\tau_+, \quad \textmd{i.e.} \quad  \tau_+= l/(c - {v}) \label{26}
\end{equation}
and for the return path
\begin{equation}
c\tau_- = l - {v}\tau_- \quad \textmd{i.e.} \quad \tau_- =l/(l+{v}).
\label{27}
\end{equation}
Summing (\ref{26}) and (\ref{27}) we find the period of oscillations
\begin{equation}
\tau = \tau_+ + \tau_- =\frac{2l/c}{1 - {v}^2/c^2}. \label{28}
\end{equation}
Substituting the formula (\ref{19}) of the Lorentz contraction in (\ref{28}) gives with the account of (\ref{6})
\begin{equation}
\tau = (2l_0/c)\gamma. \label{29}
\end{equation}
Comparing (\ref{29}) with (\ref{25}) we find with the account of (\ref{6}) that a moving clock ticks at a slower rate than the clock at  rest:
\begin{equation}
\tau = \gamma\tau_0. \label{30}
\end{equation}
If, for example, we will consider a time fuse that brings an explosive into effect after $n$ ticks then relation (\ref{30}) can be interpreted as the increase of the life-span of the object in question.

The same formula (\ref{30}) can be obtained using a transverse clock (see e.g. \cite{Dmitr}). We however succeeded in doing it in one dimension.

\section{The time is the reading of a measuring device, the
clock}

We want to measure the time interval between two events occurred in one and the same point of
the space. In the current context there can be only two kinds of events. They are the arrival and
departure of a light pulse. We will consider events that are  the beginning and the end of the
following  \emph{cyclic process}.
A clock is at $x'$ in the moving RF. There is a mirror at some distance from the clock. At
some moment of time the pulse of light is emitted from $x'$. The light reflects back from
the mirror and returns to the source.  At this moment we measure interval $\Delta t'(x')$ of
the time elapsed. Here the argument of time indicates that events occurred at
$x'$.  The procedure of measuring of  the time implies that $\Delta t'(x')$ is
in the following way related with the measurement $\Delta t$ of the same interval by a clock
at rest
\begin{equation}
\tau\Delta t'(x') = \tau_0\Delta t. \label{31}
\end{equation}
Substituting (\ref{30}) in (\ref{31}) we get
\begin{equation}
\Delta t'(x') = \Delta t/\gamma. \label{32}
\end{equation}
Relation (\ref{32}) expresses the effect of the increase of the time of a process due to the motion of the system.

\section{Measurement of the speed of light}

We are interested in the experimental determination of the speed of light in a moving RF. To this
end we will use the procedure above described. Let the reverting mirror be at the distance
$\Delta x'$ from the point  $x'$ where the clock is placed.  \textit{The speed of light is
defined as the average two-way value in the cyclic process}. Thus we find the speed of light
from the relation
\begin{equation}
c' = 2\Delta x'/\Delta t'. \label{33}
\end{equation}
Let $\Delta x$ be the distance from the source of light to the reverting mirror as measured
in ARF. We will compute the total time $\Delta t$ for the light to reach the mirror and then
to return to the source as measured in ARF. Thereto we will reproduce formulae used in (\ref{26})-(\ref{28}).
The time needed for the pulse of light to reach the mirror is
\begin{equation}
 \Delta t_+ = \Delta x/(c - {v}). \label{34}
\end{equation}
The time needed for the light to return from the mirror to the source is
\begin{equation}
\Delta t_- = \Delta x/(c + {v}). \label{35}
\end{equation}
The total time of the experiment can be found summing (\ref{34}) and (\ref{35}).
We get with the account of (\ref{6})
\begin{equation}
\Delta t = \Delta t_+ + \Delta t_- = \gamma^2\Delta x/c. \label{36}
\end{equation}
On the other side, substituting  (\ref{21}) and  (\ref{32}) into (\ref{33}) gives
\begin{equation}
c' = \gamma^2\Delta x/\Delta t. \label{37}
\end{equation}
We get comparing (\ref{37}) and (\ref{36})
\begin{equation}
c' = c. \label{38}
\end{equation}

\section{The time interval of an inertial RF between events of the absolute RF}

Now when equality (\ref{38}) of speeds of light in inertial reference frames is established we can
compute interval   $\Delta t'(x)$ of the time of IRF between two events occurred in a
given point $x$ of ARF. We will consider the following cyclic process. The source of
light is at $x$. A reverting mirror is at the point of ARF
\begin{equation}
x_+ = x + \Delta x. \label{39}
\end{equation}
At the initial instance $t = 0$ of ARF time the pulse of light is emitted from the point
$x$ of ARF in the direction of $x_+$. By the Lorentz transformation (\ref{24}) the
point $x$ of ARF has the following coordinate of IRF
\begin{equation}
x' = \gamma x. \label{40}
\end{equation}
At the moment $t = \Delta x/c$ of ARF time the light reaches the point
$x_+$ (\ref{39}) of ARF that by (\ref{24}) has
the following coordinate of IRF
\begin{equation}
x'_+ = \gamma(x_+ - {v}\Delta x/c). \label{41}
\end{equation}
After the reflection, by the moment $t = 2\Delta x/c$ of ARF time the light
returns to the point $x = x_-$ of ARF that by (\ref{24}) has the following coordinate of IRF
\begin{equation}
x'_- = \gamma(x - {v}2\Delta x/c). \label{42}
\end{equation}
The total distance passed by the light in the IRF equals to
\begin{equation}
\Delta x'_+ + \Delta x'_- = x'_+ - x' + x'_+ - x'_- = 2x'_+ - x' - x'_-. \label{43}
\end{equation}
Substituting (\ref{40})-(\ref{42}) in (\ref{43}) we get with the account of (\ref{39})
\begin{equation}
\Delta x'_+ + \Delta x'_- = 2\gamma\Delta x. \label{44}
\end{equation}
The total ARF time of this process equals to
\begin{equation}
\Delta t = 2\Delta x/c. \label{45}
\end{equation}
In a similar way as in (\ref{45}) we will find the total IRF time $\Delta t'(x)$ of this
process. We have with the account of (\ref{38})
\begin{equation}
\Delta t'(x) = (\Delta x'_+ + \Delta x'_-)/c' =
(\Delta x'_+ + \Delta x'_-)/c. \label{46}
\end{equation}
Substituting (\ref{44}) in (\ref{46}) gives
\begin{equation}
\Delta t'(x) = 2\gamma\Delta x/c. \label{47}
\end{equation}
Comparing (\ref{47}) and (\ref{45}) we find the connection of ARF time interval
$\Delta t$ with IRF time interval $\Delta t'(x)$ that elapsed between two events
occurred in the point $x$ of ARF:
\begin{equation}
\Delta t'(x) = \gamma\Delta t. \label{48}
\end{equation}
Relation (\ref{48}) will be used in derivation of the transformation of time
 to a moving RF. It is instructive to compare (\ref{48}) with the relation  (\ref{32})
obtained earlier.

\section{Synchronization of clocks}
Being  in ARF we will set spaced over $x$ clocks to one and the same time. Let the light
clock placed in the origin of coordinates ticked off  $n$  periods. The time shown by this
timepiece will be
\begin{equation}
\bar{t} = n\tau_0. \label{49}
\end{equation}
Substituting (\ref{25}) in (\ref{49}):
\begin{equation}
\bar{t} = n(2l_0/c). \label{50}
\end{equation}
By the construction of the light clock-work, during this time (\ref{50}) the light inside
the clock have covered the distance $2l_0n$. The pulse of light
emitted from the origin of coordinates at the beginning of the reading of time
will cover the same distance and reach the point
\begin{equation}
x = 2l_0n. \label{51}
\end{equation}
Substituting (\ref{51}) in (\ref{50}) we find
\begin{equation}
\bar{t} = x/c. \label{52}
\end{equation}
Thus, for synchronization each clock in ARF must be set to the time (\ref{52}) at the instant
 when the light signal emitted at the beginning of the time recording arrives at the point $x$
  of the clock's location.

From the absolute point of view in a moving reference frame the speed of light depends on the
direction of the light's motion (see  (\ref{34}) and (\ref35)). However, being in the
moving RF, we do not know about this and synchronize spaced clocks by means of the same procedure
as was used for the synchronization of clocks in ARF. We set the clock at $x'$ to the time
\begin{equation}
\bar{t}' = x'/c \label{53}
\end{equation}
at the moment when the light signal, emitted from the origin of coordinates at the beginning
 of time, comes to the point $x'$. Formula (\ref{53}) is analogous to (\ref{52}). In accord
 with (\ref{33}),  we suppose in (\ref{53}) that the speed of light is one and the same in all
directions and has the same value (\ref{38}) as before.

\section{Transformation of time}

Let us synchronize simultaneously  clocks in the rest RF and clocks in the moving RF.
We will set the reading of the clock located at the point $x$ of ARF by means of
the formula (\ref{52}). Let origins of IRF and ARF coincide at zeroth moment of time.
Then reference frames are related with each other by the transformation (\ref{24})
so that at the moment $\bar{t}$ of the absolute time $t$ the moving clock will have
 the space coordinate in IRF
\begin{equation}
x' =\gamma(x - {v}\bar{t}). \label{54}
\end{equation}
Substituting (\ref{54}) into (\ref{53}) we find the reading which the moving clock occurred
at this moment of time in $x$ should be set to:
\begin{equation}
\bar{t}' = \gamma(x - {v}\bar{t})/c. \label{55}
\end{equation}
Substituting (\ref{52}) into (\ref{55}):
\begin{equation}
\bar{t}' = \gamma(1 - {v}/c)x/c. \label{56}
\end{equation}
Formulae (\ref{52}) and (\ref{56}) are used to carry out a concerted synchronization of
clocks in inertial reference frames.

We want to describe in absolute terms the further run of IRF time $t'$ at the point
$x$ of ARF. Time intervals $t' - \bar{t}'$ and $t - \bar{t}$ at $x$ are related with each other by
equation (\ref{48})
\begin{equation}
t' = \bar{t}' + \gamma(t - \bar{t}). \label{57}
\end{equation}
Substituting (\ref{52}) and (\ref{56}) into (\ref{57}) we find
\begin{equation}
t' = \gamma(t - {v}x/c^2). \label{58}
\end{equation}
Formulae  (\ref{24}) and (\ref{58}) with (\ref{6}) define the transition from ARF to a moving RF according
to:

1) the real length contraction (\ref{19}),

2)  the change of the rhythm of time (\ref{30}) following from the independence
 of the light speed on the source and (\ref{19}),

3) the invariance of the light speed, defined as a round-trip value, that follows from the Lorentz contraction (\ref{19}) and time dilation (\ref{32}),

4)  the dependence of the time reference point on the space coordinate that follows from the isotropy of the space in the moving RF assumed by us conditionally (and unconsciously)
 in  (\ref{53}).

\section{The Lorentz group}

A uniformly moving along a straight line reference frame was shown above to be related with the
absolute reference frame by the Lorentz transformation (\ref{24}), (\ref{58}).
We will show now that the transition between two IRFs
\begin{equation}
\textmd{IRF}_1 \rightarrow \textmd{IRF}_2 \label{59}
\end{equation}
is carried out by a Lorentz transformation too. Let us accomplish
(\ref{59}) in the following way
\begin{eqnarray}
\textmd{IRF}_1 &\rightarrow & \textmd{ARF}, \label{60}\\
\textmd{ARF} &\rightarrow & \textmd{IRF}_2. \label{61}
\end{eqnarray}
Transition ARF $\rightarrow$ IRF is given by the transformation (\ref{24}),
 (\ref{58}). The inverse transformation IRF $\rightarrow$ ARF can be found from (\ref{24}), (\ref{58})
  with the account of (\ref{6}):
\begin{eqnarray}
x &=& \gamma (x' + {v}t'), \label{62}\\
t &=& \gamma (t' + x'{v}/c^2). \label{63}
\end{eqnarray}
Let $\textmd{IRF}_1$ move relative to $\textmd{ARF}$ with the
velocity ${v}_1$.  From (\ref{62}), (\ref{63}) we have for the transition (\ref{60}) with the
account of (\ref{6})
\begin{eqnarray}
x &=& \frac{x_1 + {v}_1t_1}{\sqrt{1 - {{v}_1}^2/c^2}}, \label{64}\\
t &=& \frac{t_1 + x_1 {v}_1/c^2}{\sqrt{1 - {{v}_1}^2/c^2}}. \label{65}
\end{eqnarray}
Let $\textmd{IRF}_2$ move relative to $\textmd{ARF}$ with the
velocity ${v}_2$. From (\ref{24}), (\ref{58}) we have for
the transition (\ref{61}) with the account (\ref{6})
\begin{eqnarray}
x_2 &=& \frac{x - {v}_2t}{\sqrt{1 - {{v}_2}^2/c^2}}, \label{66}\\
t_2 &=& \frac{t - x{v}_2/c^2}{\sqrt{1 - {{v}_2}^2/c^2}}. \label{67}
\end{eqnarray}
Substituting (\ref{64}), (\ref{65}) into (\ref{66}), (\ref{67}) we will express
 $x_2$, $t_2$ through $x_1$, $t_1$:
\begin{eqnarray}
x_2 &=& \frac{x_1 - {v}_{12}t_1}{\sqrt{1 - {{v}_{12}}^2/c^2}}, \label{68}\\
t_2 &=& \frac{t_1 - x_1{v}_{12}/c^2}{\sqrt{1 - {{v}_{12}}^2/c^2}} \label{69}
\end{eqnarray}
where
\begin{equation}
{v}_{12} = \frac{{v}_2 - {v}_1}{1 - {v}_1{v}_2/c^2}. \label{70}
\end{equation}
According to (\ref{68}), (\ref{69}) any IRFs are related with each other by a Lorentz
 transformation. Besides we proved that Lorentz transformations form a group with
  the group operation (\ref{70}), here given for parameters ${v}_2$ and $-{v}_1$. The transformation group is the way that the transition between IRFs can be parameterized.

A question remains. Can we consider  ${v}_{12}$ as a motion velocity
of $\textmd{IRF}_2$ relative to $\textmd{IRF}_1$?

\section{The principle of relativity}

We will show that in IRF Maxwell's equations have the same form as in ARF. Using (\ref{2})
 and (\ref{11}) in (\ref{14}) we obtain for the charge $q$ that moves
uniformly in ARF with the velocity ${v}_2$
\begin{equation}
\partial_x^2E - c^{-2}\partial_t^2E =
4\pi q[\partial_x + ({v}_2/c^2)\partial_t]\delta(x - {v}_2t) \label{71}
\end{equation}
(see a note to Eq.~(\ref{1})). Let us pass over to $\textmd{IRF}_1$. The left-hand part of
equation (\ref{71}) is identical to that in the d'Alembert equation, hence
it is Lorentz-invariant. Substituting (\ref{64}), (\ref{65}) into the right-hand part
of (\ref{71}) we will get with the account of (\ref{8}) and (\ref{70})
\begin{equation}
\delta(x - {v}_2t) = (1 - {v}_1^2/c^2)^{1/2}\delta(x_1 - {v}_{12}t_1)/(1 - {v}_1{v}_2)/c^2),\label{72}
\end{equation}
\begin{equation}
[\partial_x + ({v}_2^2/c^2)\partial_t] = (1 - {v}_1{v}_2/c^2)[\partial_{x_1} +
({v}_{12}/c^2)\partial_{t_1}]/(1 - {v}_1^2/c^2)^{1/2}. \label{73}
\end{equation}
Using (\ref{64}), (\ref{65}) in the left-hand part of (\ref{71}) and substituting
 (\ref{72}), (\ref{73}) into the right-hand part of (\ref{71}):
 \begin{equation}
\partial_{x_1}^2E - c^{-2}\partial_{t_1}^2E = 4\pi q[\partial_{x_1} + ({v}_{12}/c^2)
\partial_{t_1}]\delta(x_1 - {v}_{12}t_1). \label{74}
\end{equation}
Equations (\ref{71}) and (\ref{74}) have one and the same form. Thus,  in one dimension
 and for electromagnetism, we proved the principle of relativity: a physical phenomenon
is described in all IRFs by the same equations.

The form of the right-hand part of equation (\ref{74}) says that ${v}_{12}$ is
the velocity of motion of the charge relative to $\textmd{IRF}_1$.  Insofar as  the charge
 is at rest in $\textmd{IRF}_2$ this implies that ${v}_{12}$ is the motion velocity
 of $\textmd{IRF}_2$ relative to $\textmd{IRF}_1$.

The principle of relativity  tells us that in the Lorentz group no one reference frame is  preferred among others. In particular, the absolute reference  frame is indistinguishable among other inertial frames of reference.

\section{Once again on the Lorentz contraction}
Let the force field of the electric charge at rest in ARF be described by the function
\begin{equation}
F(x). \label{75}
\end{equation}
We consider the portion of the rest field (\ref{75}) from
\begin{equation}
F_a = F(a) \label{76}
\end{equation}
to
\begin{equation}
F_b = F(b). \label{77}
\end{equation}
The distance
\begin{equation}
l_0 = b - a \label{78}
\end{equation}
can be interpreted as the length of the body at rest.

We showed in $\S 2$ that if the charge moves with the velocity ${v}$ relative to ARF
 its force field becomes
\begin{equation}
 F(\gamma(x - {v}t)) \label{79}
\end{equation}
where $\gamma$ is given by (\ref{6}). The boundaries $\tilde{a}$ and $\tilde{b}$
 of the moving body as measured in ARF can be found from (\ref{79}), (\ref{76}):
\begin{equation}
 a = \gamma(\tilde{a} - {v}t), \quad \textmd{i.e.} \quad \tilde{a} = \frac{a}{\gamma} + {v}t \label{80}
\end{equation}
and from (\ref{79}), (\ref{77}):
\begin{equation}
 b = \gamma(\tilde{b} - {v}t), \quad \textmd{i.e.} \quad \tilde{b} = \frac{b}{\gamma} + {v}t. \label{81}
\end{equation}
Substituting (\ref{80}) and (\ref{81}) into
\begin{equation}
l = \tilde{b}-\tilde{a}\label{82}
\end{equation}
we find the length of the moving body in units of ARF:
\begin{equation}
l = \frac{b - a}{\gamma} = \frac{l_0}{\gamma}\label{83}
\end{equation}
where (\ref{78}) was used.

In $\S 3-\S 9$ we showed that the moving IRF
 is related with ARF by the Lorentz transformation  (\ref{24}),  (\ref{58}). There was
 established in $\S 11$ that Maxwell's equations are invariant under the Lorentz transformation.
 This, in particular, implies that the force field of the charge, which is at rest
in the moving IRF, is described by the function
 \begin{equation}
 F(x').\label{84}
\end{equation}
Through (\ref{84}) and (\ref{76}), (\ref{77}) the length of the moving body as viewed
 from the moving IRF is given by
\begin{equation}
l' = b - a. \label{85}
\end{equation}
Using (\ref{78}) and (\ref{83}) in (\ref{85}) we get
\begin{equation}
l' = l_0 = \gamma l. \label{86}
\end{equation}

Solving Maxwell's equations in the moving IRF we find by analogy with (\ref{79}) that the force field of the charge
moving in this IRF with the velocity $-\upsilon$ is
 \begin{equation}
F(\gamma(x' + {v}t')). \label{87}
\end{equation}
The boundaries of the body at rest in ARF as viewed from the moving IRF are given from
 (\ref{87}), (\ref{76})
\begin{equation}
 a = \gamma(a' + {v}t'), \quad \textmd{i.e.} \quad a' = \frac{a}{\gamma} - {v}t' \label{88}
\end{equation}
and from (\ref{87}), (\ref{77}):
\begin{equation}
 b = \gamma(b' - {v}t'), \quad \textmd{i.e.} \quad b' = \frac{b}{\gamma} - {v}t'. \label{89}
\end{equation}
From (\ref{88}) and (\ref{89}) we get for the length of the body at rest in ARF
 as viewed from the moving IRF
\begin{equation}
l'_0 = b' - a' = \frac{b - a}{\gamma}.\label{90}
\end{equation}
Using (\ref{85}) in (\ref{90}) we find
\begin{equation}
l'_0 = \frac{l'}{\gamma}. \label{91}
\end{equation}
Canceling primes in (\ref{91}) we may obtain a (wrong) formula that contradicts (\ref{83}),
or (\ref{19}). This will be just a supposed paradox of the Lorentz contraction.

At last, we will write down a formula of the change of the real length of a body for two moving IRFs
(\ref{59}). By (\ref{19}), (\ref{6}) we have for a body that moves
with the velocity ${v}_1$ relative to ARF
\begin{equation}
l_1 = l_0\sqrt{1 - {v}_1^2/c^2}. \label{92}
\end{equation}
We have for a body that moves relative to ARF with the velocity ${v}_2$
\begin{equation}
l_2 = l_0\sqrt{1 - {v}_2^2/c^2}. \label{93}
\end{equation}
From (\ref{92}) and (\ref{93}) we find
\begin{equation}
l_2 = l_1\frac{\sqrt{1 - {v}_2^2/c^2}}{\sqrt{1 - {v}_1^2/c^2}}. \label{94}
\end{equation}
In this event, obviously
\begin{equation}
l_2 \ne l_1\sqrt{1 - {v}_{12}^2/c^2} \label{95}
\end{equation}
where the velocity ${v}_{12}$ of $\textmd{IRF}_2$ relative to
$\textmd{IRF}_1$  is given by the formula (\ref{70}).

\section{Conclusion}

In the conclusion I shall give a brief account of basic concepts and successive steps made in order to construct special relativity theory (SRT) from the absolute reference frame (ARF).

SRT is a phenomenological theory that describes the most general properties of vacuum referred to as the symmetry. In the current context vacuum corresponds to the absolute reference frame. Formally the relativistic symmetry of vacuum is described by a one-parameter transformation of space and time coordinates which leaves invariant laws of physical phenomena occurred in the vacuum. Empirically the symmetry transformation corresponds to the transition between inertial reference frames (IRFs). The velocity ${v}$ of the relative motion of IRFs stands for the parameter of the transformation. SRT just consists in that we derive the transformation of space and time coordinates from  \textit{experimental measurements of lengths and time
intervals}.

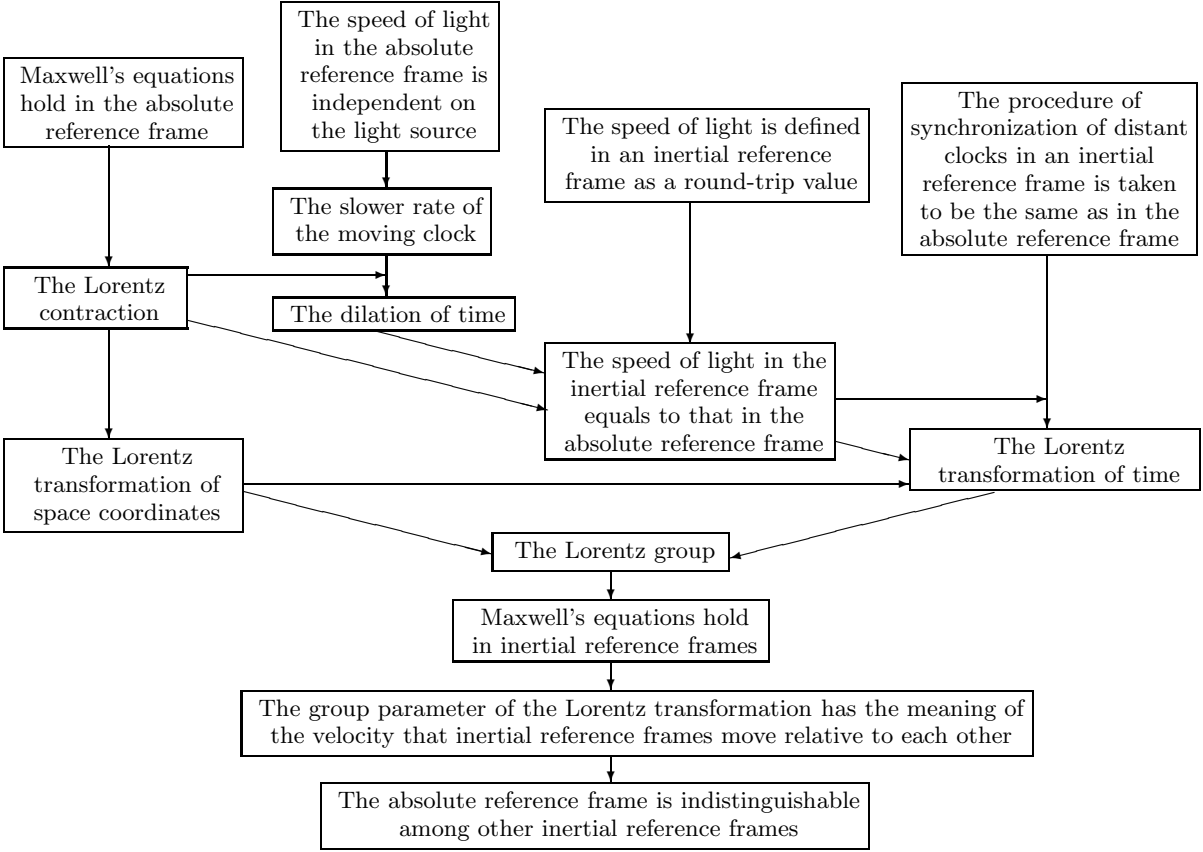
\begin{figure}[h]
\begin{picture}(500,330)(0,0)

\put(10,290){\fbox{
\begin{minipage}[]{81pt}
 Maxwell's equations hold in the absolute reference frame
 \end{minipage}}}
  \put(50,276){\vector(0,-1){46}}
\put(10,216){\fbox{
\begin{minipage}[]{60pt}
The Lorentz contraction
 \end{minipage}}}
\put(80,227){\vector(1,0){75}}
\put(80,210){\vector(4,-1){136}}
\put(50,206){\vector(0,-1){41}}
\put(10,145){\fbox{
\begin{minipage}[]{81pt}
The Lorentz transformation of space coordinates
 \end{minipage}}}
\put(101,148){\vector(1,0){252}}
 \put(101,145){\vector(4,-1){94}}

\put(115,300){\fbox{
\begin{minipage}[]{73pt}
 The speed of light in the absolute reference frame is independent on the light source
 \end{minipage}}}
 \put(155,274){\vector(0,-1){14}}
 \put(112,245){\fbox{
\begin{minipage}[]{73pt}
 The slower rate of the moving clock
 \end{minipage}}}
 \put(155,235){\vector(0,-1){16}}
  \put(112,210){\fbox{
\begin{minipage}[]{82pt}
 The dilation of time
 \end{minipage}}}
 \put(150,206){\vector(4,-1){65}}

\put(215,270){\fbox{
\begin{minipage}[]{113pt}
 The speed of light is defined in an inertial reference frame as a round-trip value
 \end{minipage}}}
 \put(270,254){\vector(0,-1){53}}
 \put(215,177){\fbox{
\begin{minipage}[]{100pt}
 The speed of light in the inertial reference frame equals to that in the absolute reference frame
 \end{minipage}}}
\put(325,180){\vector(1,0){80}}
 \put(325,164){\vector(4,-1){28}}

 \put(350,265){\fbox{\begin{minipage}[]{105pt}
 The procedure of synchronization of distant clocks in an inertial reference frame is taken to be the same as in the absolute reference frame
 \end{minipage}}}
 \put(405, 235){\vector(0,-1){66}}
 \put(353,155){\fbox{
\begin{minipage}[]{100pt}
The Lorentz transformation of time
 \end{minipage}}}
\put(385,145){\vector(-4,-1){100}}

  \put(195,120){\fbox{
\begin{minipage}[]{80pt}
 The Lorentz group
 \end{minipage}}}
 \put(240,115){\vector(0,-1){11}}
   \put(180,90){\fbox{
\begin{minipage}[]{110pt}
 Maxwell's equations hold in inertial reference frames
 \end{minipage}}}
  \put(240,80){\vector(0,-1){10}}
   \put(100,55){\fbox{
\begin{minipage}[]{290pt}
 The group parameter of the Lorentz transformation has the meaning of the velocity that inertial reference frames move relative to each other
 \end{minipage}}}
  \put(240,45){\vector(0,-1){10}}
    \put(130,20){\fbox{
\begin{minipage}[]{198pt}
 The absolute reference frame is indistinguishable among other inertial reference frames
 \end{minipage}}}

\end{picture}
\caption{\label{fig1} The "principle" of relativity derived for electromagnetism.}
\end{figure}

\begin{center}
Initial assumptions of SRT are laws that render principal properties of
vacuum.
\end{center}
\begin{description}
\item [i] Maxwell's equations are valid in ARF.
\item [ii] The speed of light does not depend on the motion of a source of the light.
\end{description}
\begin{center}
Successive derivation of relativistic transformations of the space coordinate
and time
\end{center}
\begin{description}
\item [1] Lorentz contraction (\ref{19}), (\ref{6}) is the shortening of a
moving body relative to ARF. It is a real effect that can be obtained solving Maxwell's equation
or taken for granted instead of the postulate $\textbf{i}$.
\item [2] Transformation of the space coordinate (\ref{24}) can be found from the
measurement of lengths in accord with $\textbf{1}$ and combining it with the Galilean transformation.
\item [3] When a clock moves its period increases by (\ref{30}). This is a real effect
implied by the independence of the light speed on the motion of the
source with the account of the Lorentz contraction.
\item [4] The decrease in the time of the cyclic process of IRF in a fixed point of IRF (\ref{32}) can be found from the measurement of the time interval according to $\textbf{3}$.
\item [5] The invariance of the speed of light (\ref{38}) can be found computing the
speed of light as the average value in the cyclic process and using $\textbf{4}$.
\item [6] We calculate the time of the cyclic process in IRF (\ref{48}) at a fixed
point of ARF  using the invariance of the speed of light proved in $\textbf{5}$.
\item [7] The synchronization of clocks in IRF was made using the isotropic formula (\ref{53})
 where $\textbf{5}$ was taken into account.
\item [8] The transformation of time  (\ref{58}) can be derived using $\textbf{7}$ and $\textbf{6}$.
\end{description}
See the scheme in Fig.\ref{fig1}.

\appendix
\section{The change in dimensions of a moving body}
\label{contraction}
We will assume that the interaction between atoms has the electromagnetic origin, i.e. it can be reduced ultimately to the force between electric charges. Consider two charges $q$ and $q_0$ in ARF at  $(0,0,0)$ and $(x,y,z)$ respectively. Let the charges move uniformly along the $x$-axis. It can be shown  \cite{Dmitriyev} that the force of interaction depends on the distance between the charges and velocity ${v}$ of the motion in the following way:
\begin{equation}
{\bf F} =
q_0q\frac{\gamma(x - {v}t){\bf i}+(y{\bf j}+z{\bf k})/\gamma}{[\gamma^2(x - {v}t)^2 + y^2 + z^2]^{3/2}} \label{c1}
\end{equation}
where parameter $\gamma$ depends on the velocity by (\ref{6}). From (\ref{c1}) any force field can be represented as a sum of vector functions shifted and scaled along the $x$-axis:
\begin{equation}
{\bf F} =
F_x(\gamma(x - {v}t), y, z){\bf i} + \frac{1}{\gamma}[F_y(\gamma(x - {v}t), y, z){\bf j} + F_z(\gamma(x - {v}t), y, z){\bf k}]. \label{c2}
\end{equation}
In the equilibrium point ($\check{x},\check{y},\check{z}$) the interaction force (\ref{c2}) should be nullified, i.e.
\begin{eqnarray}
F_x(\gamma(\check{x} - {v}t),\check{y},\check{z}) &=& 0, \label{c3}\\
F_y(\gamma(\check{x} - {v}t),\check{y},\check{z}) &=& 0, \label{c4}\\
F_z(\gamma(\check{x} - {v}t),\check{y},\check{z}) &=& 0. \label{c5}
\end{eqnarray}
From (\ref{c3})-(\ref{c5}) and (\ref{6}) the relation between ($\check{x}, \check{y}, \check{z}$) and the equilibrium point  ($\check{x}_0, \check{y}_0, \check{z}_0)$ of the system at rest can be found:
\begin{eqnarray}
\gamma(\check{x} - {v}t) &=& \check{x}_0, \label{c6}\\
\check{y} &=& \check{y}_0, \label{c7}\\
\check{z} &=& \check{z}_0. \label{c8}
\end{eqnarray}
Relation (\ref{c6}) expresses the Lorentz-Fitzgerald contraction (\ref{19})
\begin{equation}
\check{x} - {v}t = \check{x}_0/\gamma. \label{c9}
\end{equation}
By (\ref{c7}) and (\ref{c8}) the transverse dimensions of the body stay unchanged.

\section{Relativistic mechanics}
\label{mechanics}
By the principle of relativity, equations of classical mechanics
must have one and the same form in each IRF. Let us show that this
requirement is satisfied by the form in the left-hand part of the
following motion equation
\begin{equation}
m\frac{d^2x/dt^2}{[1 - (dx/dt)^2/c^2]^{3/2}} = F. \label{a1}
\end{equation}
Taking differential of (\ref{62}) and (\ref{63})
\begin{eqnarray}
dx &= &\gamma(dx' + {v}dt'), \label{a2}\\
dt &=& \gamma(dt' + dx'{v}/c^2). \label{a3}
\end{eqnarray}
Dividing (\ref{a2}) by (\ref{a3}):
\begin{equation}
dx/dt = \frac{dx'/dt' + {v}}{1 + (dx'/dt'){v}/c^2}. \label{a4}
\end{equation}
Rewriting (\ref{a4}) into a linear form, taking its
differential and dividing by (\ref{a3}):
\begin{equation}
d^2x/dt^2 = \frac{(d^2x'/dt'^2)(1 - (dx/dt)v/c^2)}{1 + (dx'/dt')v/c^2}.
\label{a5}
\end{equation}
Substituting (\ref{a4}) into (\ref{a5}) we obtain with the
account of (\ref{6})
\begin{equation}
d^2x/dt^2 = \frac{d^2x'/dt'^2}{\gamma^2[1 + (dx'/dt'){v}/c^2]^3}.
\label{a6}
\end{equation}
We will substitute (\ref{a4}) and (\ref{a6}) into (\ref{a1})
taking into account (\ref{6}). Thus we show that the form of
the left-hand part of (\ref{a1}) is retained in the transformation
between inertial reference frames:
\begin{equation}
m\frac{d^2x/dt^2}{[1 - (dx/dt)^2/c^2]^{3/2}} = m\frac{d^2x'/dt'^2}{[1 - (dx'/dt')^2/c^2]^{3/2}}.
\label{a7}
\end{equation}
Expression (\ref{a1}) turns into that of classical mechanics when
$c\rightarrow\infty$. Equation of relativistic mechanics
(\ref{a1}) can be also rewritten in the form
\begin{equation}
\frac{d}{dt}\left\{\frac{mdx/dt}{[1 - (dx/dt)^2/c^2]^{1/2}}\right\} = F.
\label{a8}
\end{equation}

\section{A spring clockwork}
\label{pendulum}
 We will consider a linear oscillator
\begin{equation}
m\frac{d^2x}{dt^2} = -kx. \label{b1}
\end{equation}
The equation of classical mechanics (\ref{b1}) describes the
process of vibration of the mass $m$ under the action of the
restoring force. Solving (\ref{b1}) we find the period of this
process
\begin{equation}
\tau_0 = \sqrt{2\pi m/k}. \label{b2}
\end{equation}
The duration (\ref{b2}) can be taken as a unit of time.

If the oscillator moves as a whole with the velocity ${v}$ we must
introduce in the expression of the force the displacement from the
equilibrium point ${v}t$ and also the scaling factor $\gamma$
(\ref{6}), i.e. in accord with (\ref{18}) the force will be
\begin{equation}
-\gamma k(x - {v}t). \label{b3}
\end{equation}
Substituting (\ref{b3}) into relativistic equation of motion
(\ref{a1}):
\begin{equation}
m\frac{d^2x/dt^2}{[1 - (dx/dt)^2/c^2]^{3/2}} = -\gamma k(x - {v}t).
\label{b4}
\end{equation}
We will solve equation (\ref{b4}) provided that the velocity of
oscillations is much less than the velocity of the translational
motion of the oscillator:
\begin{equation}
|dx/dt - {v}|\ll{v} . \label{b5}
\end{equation}
Then with the account of (\ref{6}) equation (\ref{b4}) will be
\begin{equation}
\gamma^2md^2x/dt^2 = -k(x - {v}t). \label{b6}
\end{equation}
From (\ref{b6}) we obtain the period of vibration of the moving
oscillator
\begin{equation}
\tau = \gamma\sqrt{2\pi m/k}. \label{b7}
\end{equation}
Comparing (\ref{b7}) and (\ref{b2}) we find with the account of
(\ref{6}) that a moving clock shows the dilation of time
(\ref{30}).

\end{document}